\documentclass[sigconf]{acmart}
\AtBeginDocument{%
  }

\setcopyright{acmcopyright}
\copyrightyear{2022} 
\acmYear{2022} 
\setcopyright{rightsretained} 
\acmConference[AutomotiveUI '22]{14th International Conference on Automotive User Interfaces and Interactive Vehicular Applications}{September 17--20, 2022}{Seoul, Republic of Korea}
\acmBooktitle{14th International Conference on Automotive User Interfaces and Interactive Vehicular Applications (AutomotiveUI '22), September 17--20, 2022, Seoul, Republic of Korea}\acmDOI{10.1145/3543174.3546835}
\acmISBN{978-1-4503-9415-4/22/09}
\usepackage{graphicx}
\usepackage{subfig}
\usepackage{times}
\usepackage{epsfig}
\usepackage{amsmath}

\usepackage{amssymb}
\usepackage{epsfig}
\usepackage[super]{nth}
\usepackage{tabulary}
\usepackage{caption}

\newcommand{\figref}[1]{Figure~\ref{#1}}
\newcommand{\tabref}[1]{Table~\ref{#1}}

\begin{document}

%%
%% The "title" command has an optional parameter,
%% allowing the author to define a "short title" to be used in page headers.
\title{Enjoy the Ride Consciously with CAWA: Context-Aware Advisory Warnings for Automated Driving}
\renewcommand{\shorttitle}{Enjoy the Ride Consciously with CAWA}

\author{Erfan Pakdamanian}
\affiliation{
  \institution{School of Engineering\\ University of Virginia}
  \city{Charlottesville, VA}
  \country{USA}
}
\email{ep2ca@virginia.edu}

\author{Erzhen Hu}
\affiliation{
  \institution{School of Engineering\\ University of Virginia}
  \city{Charlottesville, VA}
  \country{USA}
}
\email{eh2qs@virginia.edu}

\author{Shili Sheng}
\affiliation{
  \institution{School of Engineering\\ University of Virginia}
  \city{Charlottesville, VA}
  \country{USA}
}
\email{ss7dr@virginia.edu}

\author{Sarit Kraus}
\affiliation{
  \institution{Department of Computer Science\\ Bar-Ilan University}
  \city{Tel Aviv}
  \country{Isreal}
}
\email{sarit@cs.biu.ac.il}

\author{Seongkook Heo}
\affiliation{
  \institution{School of Engineering\\ University of Virginia}
  \city{Charlottesville, VA}
  \country{USA}
}
\email{seongkook@virginia.edu}

\author{Lu Feng}
\affiliation{
  \institution{School of Engineering\\ University of Virginia}
  \city{Charlottesville, VA}
  \country{USA}
}
\email{lu.feng@virginia.edu}

\renewcommand{\shortauthors}{Pakdamanian et al.}

%====================================
\begin{abstract}
In conditionally automated driving, drivers decoupled from driving while immersed in non-driving-related tasks (NDRTs) could potentially either miss the system-initiated takeover request (TOR) or a sudden TOR may startle them. To better prepare drivers for a safer takeover in an emergency, we propose novel context-aware advisory warnings (CAWA) for automated driving to gently inform drivers. This will help them stay vigilant while engaging in NDRTs. The key innovation is that CAWA adapts warning modalities according to the context of NDRTs. We conducted a user study to investigate the effectiveness of CAWA. The study results show that CAWA has statistically significant effects on safer takeover behavior, improved driver situational awareness, less attention demand, and more positive user feedback, compared with uniformly distributed speech-based warnings across all NDRTs.
\end{abstract}

%====================================


%====================================
%% Keywords

\begin{teaserfigure}
\centering
   \includegraphics[width=0.73\textwidth]{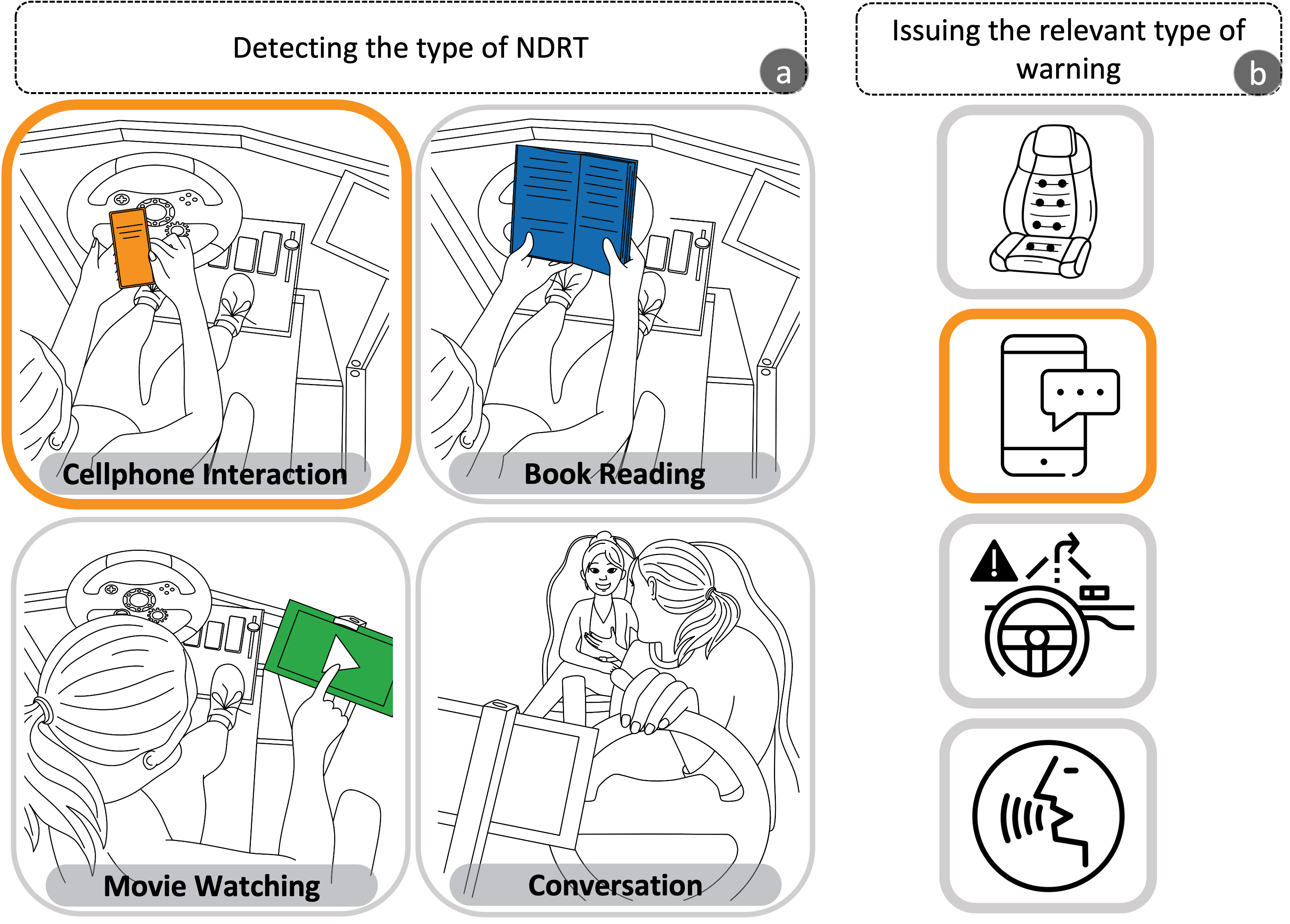}
   \caption{The study's proposed context-aware advisory warning method, \textit{CAWA}. a) Detection of the NDRT in which the driver is engaged, b) Selecting the type of modality according to detected activity.} % add highway view
   \label{fig:Teaser}
\end{teaserfigure}

\maketitle

%====================================
%% Main text

\section{Introduction}\label{sec:intro}
The rapid development of autonomous driving technologies promises a future where drivers can take their hands off the steering wheels, foot off the pedals, and instead engage in non-driving related tasks (NDRTs) such as reading or using mobile devices. While full self-driving vehicles are not yet commercially available, we are at the stage that conditionally automated driving~(level 3 of autonomy, defined by the Society of Automotive Engineers (SAE)~\cite{sae2018taxonomy}) provides various forms of driver assistance, advanced monitoring systems, and control of the longitudinal and lateral vehicle kinematics on a sustained basis.
% require human driver to relinquish the control in case of system failures or system limited capabilities. 
Although in conditionally automated driving, drivers do not need to continuously monitor the driving environment, due to current technology limitations and legal restrictions, the automated system still needs to relinquish the control back and ask the human driver to resume the control in case of system failures, anticipated dangerous situation, or exceeding its operational limit via a so-called take-over request~(TOR)~\cite{bazilinskyy2018take, gold2013take}.
% (e.g. tragic accidents caused by automation system detection failures~\textcolor{red}{[uber and tesla]}). 

A growing body of research shows that being immersed in NDRTs for an extended period of time causes the level of situation awareness to fall below a comfortable point to safely recover manual control, mainly in urgent situations~\cite{weaver2020systematic,pakdamanian2021deeptake,marberger2017understanding}. Importantly, the control transition process and taking control back cause longer reconfiguration of cognitive and motoric states for drivers to react properly~\cite{marberger2017understanding,kerschbaum2015transforming}. Thus, human factors researchers argue while most vehicles are not completely self-driving, safety hurdles arise in automated vehicles. Recent fatal crashes indicate drivers' failures to promptly and properly respond to a TOR due to the loss of situation awareness~\cite{board2020collision}. Hence, a key challenge is how to maintain driver readiness for a safe takeover while enabling an enjoyable user experience of engaging in NDRTs. 
Most existing works focus on the design of TORs, such as its timing~\cite{yoon2021modeling, du2020evaluating} and modalities~\cite{yoon2019effects,salminen2019unimodal,pakdamanian2020toward}.
%However, TORs alone are not sufficient to guarantee safe takeover. 
On the one hand, limitations on current vehicle sensing technologies pose constraints on how early hazardous road incidents can be detected for initiating TORs. 
The takeover time-budget between the TOR initiation and the incident occurrence is typically 5-7~seconds~\cite{zhang2019determinants}, which may not be long enough for drivers immersed in NDRTs to regain situational awareness and resume manual driving in a timely and safe fashion. On the other hand, current incorporated unimodal or multimodal TOR may suddenly inform drivers about an upcoming hazard~\cite{zhang2019determinants}, which may in fact startle and stress the driver and leaving the driver in a less capable state to execute a life-saving maneuver.

%\textcolor{red}{\\emph{Advisory warnings} are issued before imminent driver warnings to direct driver attention to a potential hazard in order to increase readiness.}

% \begin{figure}[ht!]
% \vspace{-0.15in}
% \centering
%   \includegraphics[width=0.5\textwidth]{Figures/teaser_5.png}
%   \caption{The study's proposed context-aware advisory warning method, \textit{CAWA}. a) Detection of the NDRT in which the driver is engaged, b) Selecting the type of modality according to detected activity.\vspace{-0.15in} }
%   \label{fig:teasor}
% \end{figure} 
% % \vspace{-0.5cm}
To address the aforementioned limitations, we propose context-aware advisory warnings~(CAWA) for automated driving to gently and adaptively inform drivers~(see Figure~\ref{fig:Teaser}), helping them stay vigilant while engaging in NDRTs. Previous studies on advisory warnings mainly regard manual driving system settings that alert drivers prior an upcoming hazard~\cite{seeliger2014advisory,maag2015car}. In contrast, we consider advisory warnings for automated driving system to let drivers know that they are entering the incipient phase of error creation.
Then, the key contributions of CAWA are two-fold: 
(1) CAWA adapts warning modalities according to the NDRT context in which a driver is immersed, for reducing the likelihood that a warning will go unnoticed. 
(2) CAWA provides gentle warnings in contrast with sudden and startling TORs. 
For example, if a driver is playing a game on her mobile phone and is wearing headphones, CAWA sends a text message warning to the phone to grab the driver's attention, while auditory or visual warnings may be missed.

% \begin{figure}[t!]
% \vspace{-0.15in}
% \centering
%   \includegraphics[width=8cm, height=6cm]{Figures/teaser_5.png}
%   \caption{The study's proposed context-aware advisory warning method, \textit{CAWA}. a) Detection of the NDRT in which the driver is engaged, b) Selecting the type of modality according to detected activity.\vspace{-0.15in} }
%   \label{fig:teasor}
% \end{figure} 
% \vspace{-0.5cm}

% We conducted a user study structured around the following research questions:
% \vspace{-0.2cm}
% \begin{itemize}
%     \item \textbf{RQ1}: How does CAWA affect driver takeover behavior?
%     \item \textbf{RQ2}: How does CAWA affect driver situational awareness?
%     \item \textbf{RQ3}: How does CAWA affect driver stress and cognitive workload?
%     \item \textbf{RQ4}: What are driver perceptions (e.g., safety, disruptiveness, and urgency) about CAWA?
% \end{itemize}
% % We recruited 20 participants to conduct the user study on a driving simulator, where the 
In this study each participant experienced two driving scenarios, CAWA and baseline.
In the CAWA trial, advisory warnings were issued depending on the context of NDRTs (e.g., text message warning when the driver is playing a game on her mobile phone, visual warning when the driver is having a conversation)~(see Figure~\ref{fig:Teaser}).
In the baseline, however, auditory warnings were given uniformly for all NDRTs. We compared CAWA with auditory warning as these are omnidirectional and have already widely applied by auto-manufacturers.
The user study demonstrated promising results. Compared with the baseline, CAWA has statistically significant effects on safer takeover behavior, improved driver situational awareness, less attention demand for workload, and more positive driver perceptions.  

To the best of our knowledge, this is the first study on context-aware advisory warnings for automated driving. We believe that our work has the potential to provoke future HCI research on integrating advisory warnings into the design of automated vehicles, taking a step toward improving the safety and user experience of automated driving.

\section{Related Work} \label{sec:related}
Takeover performance can be explained by both reaction time and post takeover control~\cite{mcdonald2019toward,pakdamanian2021deeptake}. Despite many factors have been identified contributing to better reaction time and takeover control such as traffic density~\cite{gold2016taking} and driver cognitive state~\cite{sadeghian2018feel,van2021effect} or emotion~\cite{sanghavi2020effects}, the impact of time budget (“lead time”)~\cite{eriksson2017takeover} and TOR modality~\cite{borojeni2017comparing} have been widely studied by researchers. For example, studies show that additional second of time budget lead to increase of reaction time by on average 0.27second~\cite{zhang2019determinants,mcdonald2019toward}. If drivers are given more time to gain sufficient situation awareness, they could prepare for the upcoming transition of control. Gold et al.~\cite{gold2013take} has shown
that shorter takeover times lead to faster responses but worse maneuvers. On the other hand, a study by Merat et al.~\cite{merat2014transition} suggests 20-40second of time budget for a safe takeover to fully stabilised the vehicle after reclaiming control. As supplying such time budget may not be technologically feasible at the moment, researchers are required to study alternative approaches to enable drivers gaining enough situation awareness as a function of available time~\cite{lu2017much}.  

To improve takeover time and quality, many warning modalities have been studied such as audio~\cite{politis2015beep}, visual~\cite{kim2017you}, vibrotactile~\cite{bazilinskyy2018take} and combination of these warning modalities~\cite{baldwin2012multimodal}. Prior studies explored priming drivers before asking them to resume vehicle control. In the study by van der Heiden~\cite{van2017priming}, participants received audio warnings 20 seconds prior to TORs, which caused them to disengage from the NDRT earlier and look at the road more closely. In another study~\cite{hollander2018preparing}, participants received visual warnings indicating the remaining driving time or distance until a TOR would be issued. 
Compared with these existing works, our study employed a richer set of warning modalities including speech-based cues, visual head-up-displays, text messages, and vibrotactile cues. 

Previous research has extensively studied different modalities for in-vehicle alerts, in particular TORs. One of the most prevalent modalities is auditory cues, which can be divided into two categories: nonspeech- and speech-based. Compared with nonspeech-based auditory tones, speech-based messages offer more information and are more favorable to drivers~\cite{wu2021auditory}. 
Various representations of visual cues have been designed and utilized, such as a head-up-display~\cite{gerber2020self}, augment reality~\cite{lorenz2014designing}, and LED lights~\cite{borojeni2018reading}.
Studies also found that vibrotactile and haptic cues can effectively alert drivers~\cite{dass2013haptic,telpaz2015haptic,morrell2010design}.
Recent efforts have been increasingly focusing on multi-modal alerts where multiple modalities are triggered simultaneously~\cite{petermeijer2017driver,bazilinskyy2018take,sanghavi2021multimodal}.
While multi-modal alerts were found to be more effective (e.g., leading to shorter takeover reaction time), they were perceived as more urgent and annoying~\cite{politis2015language}.
Our study takes a different approach from these existing works by incorporating advisory warnings instead of TORs.
Moreover, in order to avoid prevalence alert fatigue, CAWA  chooses a proper advisory warning from multiple modalities according to the context of NDRTs, rather than triggering all modalities simultaneously.

\section{Method} \label{sec:study}

In this section, we describe the experimental setup, design and procedure. 
The study protocol was approved by the Institutional Review Board at University of Virginia (\#IRB-SBS 4701).

%====================================
\subsection{Participants}

We recruited a total of 20 participants (14 males; 6 females) with the age range of 18-32 years old (mean= 22.65years; SD=~4.01years).
All eligible participants had normal or corrected-to-normal vision, as well as a valid driver's license~(mean= 2.8 years, SD = 3.1 years). 
None of the participants had previous experience with automated driving or prior knowledge about the user study. 
We used 19 participants' data for the result analysis, excluding one participant due to largely missing biometric data.

%====================================
%--------------------
% \begin{figure}[t!]
% \centering
%   \includegraphics[width=\textwidth]{Figures/drivingSim_annt_2.png}
%   \caption{The driving simulator setup for the user study.} 
%   \label{fig:Sim}
%   \vspace{-0.15in}
% \end{figure} 
%---------------------------------------
% \begin{figure}[t!]
% \centering
% \vspace{-1cm}
%   \includegraphics[width=\textwidth]{Figures/drivingSim_annt_2.png}
%   \begin{minipage}{.5\textwidth}
%         \centering
%         \vspace{-0.15in}
%         \captionof{figure}{The driving simulator setup for the user study.}
%           \label{fig:Sim}
%           \end{minipage}%
%   \begin{minipage}{.45\textwidth}
%         \centering
%         \vspace{-0.15in}
%       \captionof{figure}{Estimated eye region landmarks and direction.}
%     \label{fig:eyetracker}
%           \end{minipage}
%   \vspace{-0.15in}
% \end{figure} 
% \vspace{-0.15in}

%===================================

%--------------------
\begin{figure}[t!]
\centering
   \includegraphics[width=0.98\linewidth]{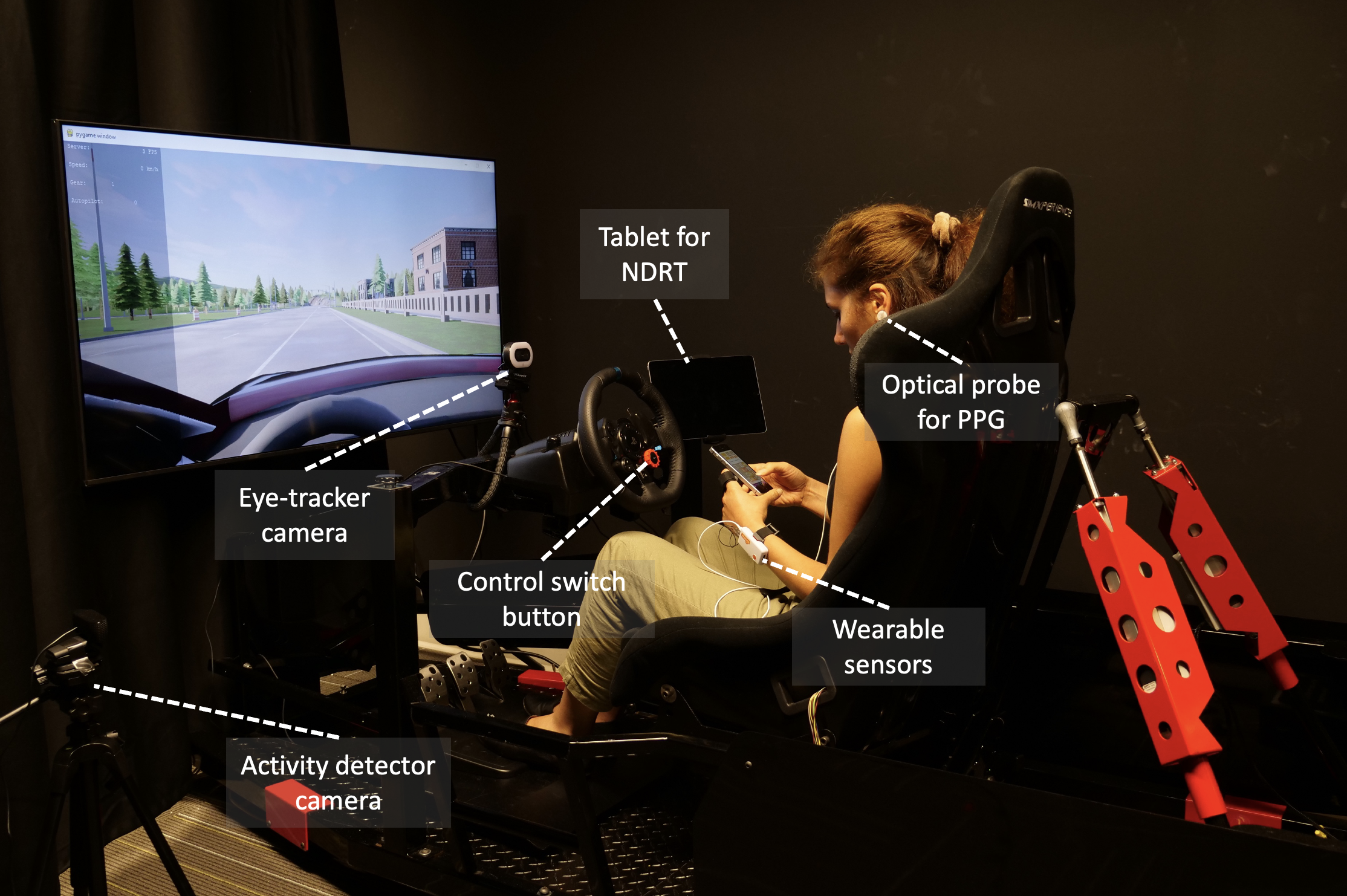}
   \caption{The driving simulator setup for the user study.}
   \label{fig:Sim}
\end{figure} 
% \vspace{-0.35in}

%--------------------
\subsection{Experimental Apparatus} 
\emph{Driving simulator}. The study was conducted in a fixed-based driving simulator from SimXperience~(Stage 5 Full Motion Racing simulator,~\figref{fig:Sim}). The setup consists of a 55-inch display~(1280 × 720 pixel resolution) placed within a horizontal field-of-view and approximately 63-inch away from the driving seat, a racing car seat, a Logitech G29 steering wheel, and sport pedals. No gearshift was required and participants could switch between automated and manual driving modes by pressing a designated button on steering wheel~(see \figref{fig:Sim} for details). An Apple iPad Pro with a 9.7-inch display was mounted on the right side of the driving seat for watching movies. Tablet was mounted in common height of the infotainment systems in a landscape format. A 2.0 channel sound bar speaker was placed behind the driver seat for the auditory warnings.
The virtual driving environment was created using CARLA~\cite{dosovitskiy2017carla}, an open-source driving simulation environment built on top of the Unreal Engine. The vehicle was programmed to simulate an SAE Level 3 automation, which handled the longitudinal and lateral vehicle kinematics, and responded to traffic elements.

\emph{Biometrics}. In this study, we collected drivers’ psychophysiological, vehicle-related metrics, workload, and perceived safety.
We used a Shimmer3+ wearable device to measure the driver's heart rate (PPG) and galvanic skin response (GSR) signals with a sampling rate of 256~Hz. Heart rate variability~(the time elapsed between two successive R-waves) from PPG and maximum and mean phasic components were calculated as the objective metrics reflecting cognitive load variation and stress, respectively. 

\begin{figure*}[t!]
\centering
   \includegraphics[width=\textwidth]{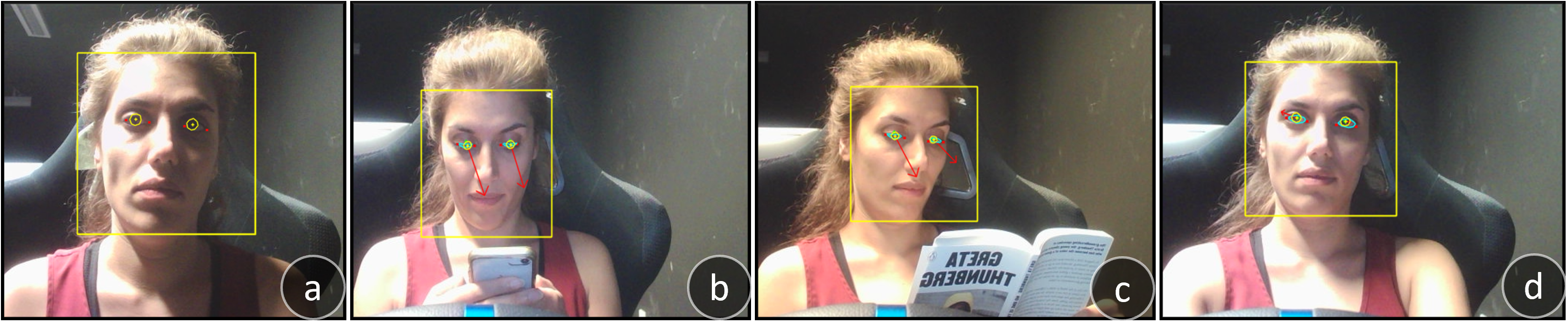}
   \caption{Examples of estimated eye region landmarks around the iris and eyelid edges along with gaze direction while performing NDRTs and after a takeover control. a) four main landmarks of eyes and pupil detection, b) gaze direction while looking at the phone, c) gaze direction while reading a book, d) looking at the road after takeover control resumption.}
   \label{fig:eyetracker}
\end{figure*}

\emph{Face and activity cameras}. We installed one high resolution camera (NexiGo N930E  1080p webcam with ring light) above the steering wheel to monitor the driver’s eye and head movements. Since CAWA required real-time detection of gaze behavior, we employed state-of-the art pupil and iris localization models~\cite{park2018learning,Xiong_2019_CVPR} and modified it to fit our needs by integrating deep pictorial gaze estimation~\cite{park2018deep,baee2021medirl}. Thus, we were able to reliably estimate position and direction of gaze in real-time. \figref{fig:eyetracker} shows an example of the face video examined to capture drivers’ eye movements and gaze directions. These videos helped to monitor and to identify when a driver detected a threat or when took her eyes off the driving scene. Furthermore, a high resolution camera (Logitech Ultra HD 1080p) was used to extract participant's driving and engagement activities. 
Finally, we developed multiple APIs to forward all stream of data to iMotions biometric platform for the real-time aggregation and synchronization. 

%--------------------
%====================================

% pPG peaks were detected using an adaptive threshold method for heart rate extraction (Shin et al., 2009). Heart rate variability was calculated as the standard deviation of RR intervals (i.e., the time elapsed between two successive R-waves on the electrocardiogram) (Castaldo et al., 2017). 
%====================================
% context-aware warnings were initiated using the Wizard-of-Oz technique where the Wizard operator issued the warnings.

\subsection{Experimental Design} 

\begin{figure}[h!]
    \centering
    \includegraphics[width=\linewidth]{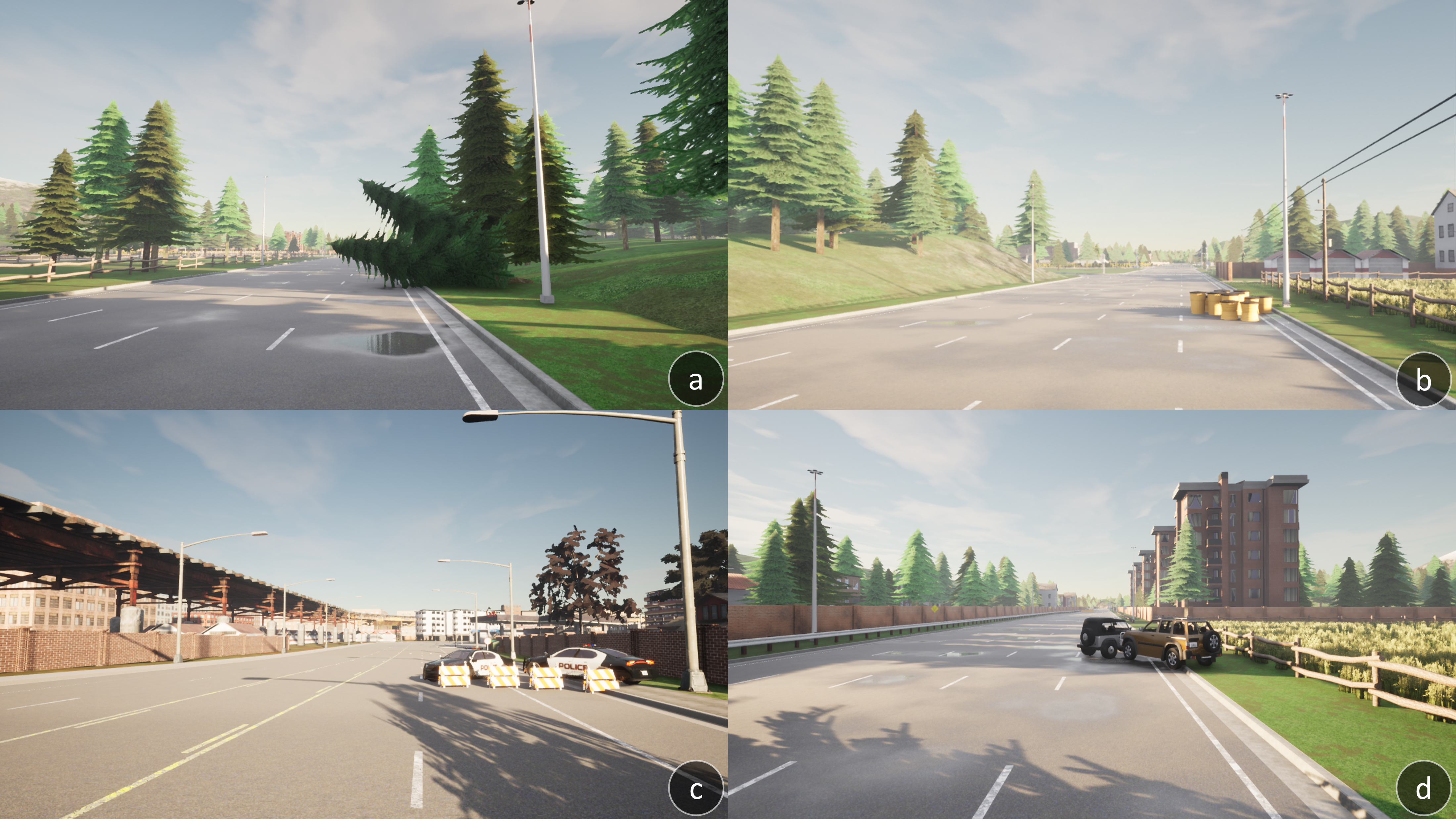}
  \caption{Examples of the TOR four takeover situations, adapted from~\cite{petermeijer2017driver,korber2018have}. a) Fallen trees. b) Working zone. c) Police set up roadblocks. d) Breakdown cars. }
  \label{fig:scenarios}
\end{figure}

We used a within-subject design with driver’s cognitive load, and the modality of advisory warnings as independent variables~(see Section \ref{sec:independent}). The cognitive load was manipulated via the difficulty of the NDRTs~(low: watching movie; mid: reading and having an informal conversation; high: playing 2048 game)~(see Table\ref{tab:NDRTs}).
These four activities were selected as the common activities drivers will most likely engage with in L3~\cite{louw2019engaging,naujoks2016secondary}. Based on prior literature~\cite{petermeijer2017driver,korber2018have}, four takeover events were designed in urban areas with typical roadway features~(see~\figref{fig:scenarios}). The difficulty of the scenarios was designed to be approximately the same. Each participant executed two sessions~(CAWA and baseline) and the order of sessions was counterbalanced across participants. Per session, the participant experienced 16 possible takeover events~(4 TORs per NDRT). In order to avoid predictably and over-trusting of the automated system, we randomly assigned 4 more TORs in each trial to be false alarms, where no hazardous incident was actually detected but a TOR was issued. Although participants interacted with all NDRTs, the given advisory warnings were different in each session. In the CAWA session, the modality of advisory warnings adapted to the context of NDRTs, whereas in baseline, all advisory warnings across different NDRTs use the same auditory modality. In both experimental sessions, the simulated vehicle was equipped with SAE Level 3 automation which could issue TORs~(350~Hz acoustic tone with 75~ms duration) to ask the driver to resume the control once it detected an unfamiliar situation out of its capabilities. In the manual driving mode, participants could control the vehicle via the steering wheel and pedals~(see details in Sec.~\ref{Sec:Procedure}).

%--------------------

%--------------------

\begin{table}[htb]
\begin{center}
\caption{CAWA adapts advisory warning modalities based on the context of NDRTs} 
\label{tab:NDRTs} 
    \begin{tabular}{c|c}
        \toprule 
        \textbf{Non-Driving Related Tasks}  & \textbf{Warning Modalities} \\
        \hline
        \hline
        Playing 2048 game on the cellphone & Text message \\
        \hline
        Watching a movie on the tablet & Vibrotactile \\
        \hline
        Reading a book & Speech-based \\
        \hline
        Conversation with the passenger & Visual\\
        \bottomrule
    \end{tabular}
\end{center}    
\end{table}
% \vspace{-5cm}

%--------------------

\subsection{Independent Variables} \label{sec:independent}
\subsubsection{Modalities}
\begin{figure*}[t!]
\centering
  \includegraphics[width=0.98\textwidth]{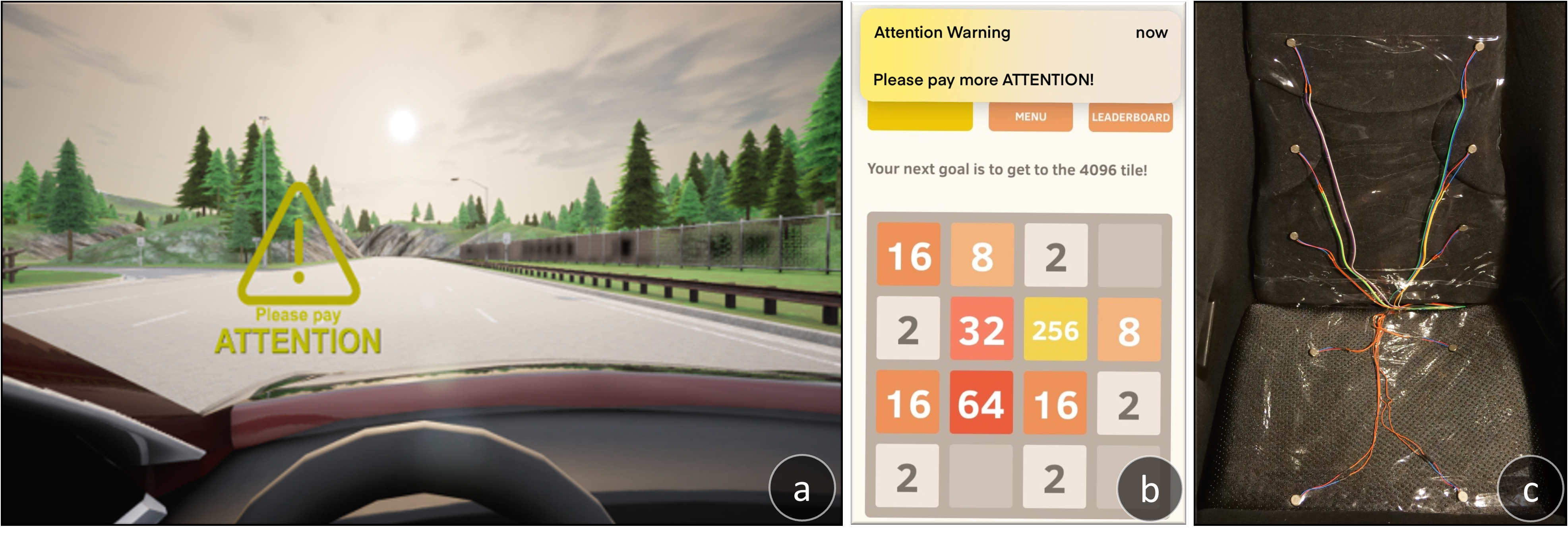}
  \caption{Advisory warning modalities: (a) visual warning from the ego's vehicle view , (b) text message, (c) vibrotactile.}
  \label{fig:modalities}
\end{figure*}

\emph{\textbf{Text message}}. We developed a Python API that can automatically send a text message containing an advisory warning of ``Please pay ATTENTION!'' to the driver's mobile phone (see \figref{fig:modalities}(b)). The developed attention warning message was displayed at the top of the screen. While drivers are immersed with playing a game on phone, they may potentially miss the auditory and visual cues. In such situation, a notification that grabs users' attention with a quick-to-the-point warning could abruptly direct their attention to the driving scene. 
    
\emph{\textbf{Vibrotactile}}. We attached 10 vibrotactile actuators (Tatoko 10mm~ \\$\times$~3mm vibration motor, 3V, 12000rpm) to the driver's seat as shown in \figref{fig:modalities}(c), and used an Arduino Uno microcontroller and L9910 motor drivers to drive the vibrotacile actuators. The generated vibrotactile feedback pattern involves two 200~ms long vibrations at maximum amplitude, separated by a 200~ms delay between them. 

\emph{\textbf{Speech-based}}. Previous research has shown that semantics and emotional tone leads to higher perceived urgency~\cite{baldwin2011verbal,politis2015language,ljungberg2012listen}. So, it is important to consider whether the message is comprehensible and pleasant for a driver to react upon in a timely manner. We created a gentle warning message ``Please pay attention'' with a female voice and an American accent. 
% An auditory speech-based warning is commonly used for conveying various information by car manufacturers. 

\emph{\textbf{Visual}}. Head-up-displays are increasingly used for effective visual communication with drivers~\cite{doshi2008novel}. We designed the visual advisory warning as a windshield projected head-up-display shown in \figref{fig:modalities}(a), which includes a warning sign icon accompanying the text ``Please pay ATTENTION''.

Please note that we implemented a unimodal advisory warning in CAWA to be effective for each NDRT and to avoid resource sharing conflicts defined by Wickens’ multiple resource theory~\cite{wickens2002multiple}.
% Additionally, the present study employed the Wizard-of-Oz protocol to issue the warning modalities based on the engaged NDRTs~(details in Sec.~\ref{Sec:Procedure}). 

% We implemented the proposed CAWA method by choosing the most appropriate and effective warning modality for each ND by intuition (see \tabref{tab:NDRTs}). 
% \ez{maybe we can use what Reviewer 1 suggest here: justify the selection use Wickens’ Multiple Resource Theory~\cite{wickens1991processing}.}
% \ez{May be we can have one more column in Table 1 addressing what demand these NDRTs require respectively so that we instead choose a different modality as an advisory warning according to this ... theory.}
%
%--------------------

%--------------------

\subsubsection{Non-driving activities}
% Humans are prone to distraction. Even in manual driving where drivers are prohibited from using hand-held devices, NDRTs prevalence of more than 50\% of the driving time for the United States~\cite{dingus2016driver}. Thus, as automated driving lets drivers to engage to unlimited tasks, drivers are more likely to increase their engagement in NDRTs. In this project, we select the most common activities that not only drivers will be likely to engaged with~\cite{yoon2019non}, but also their impact on drivers motor readiness, cognitive resource utilization, and gaze behavior~\cite{yoon2019effects}. 
Participants were asked to perform four NDRTs with three cognitive difficulty levels~(i.e. Low: watching movies; Mid:~reading and informal conversation; High:~playing a mentally demanding game) while setting the vehicle in an automated driving mode. They were also informed that they needed to take control of the vehicle in case a TOR is issued. Studies have shown that engaging with a NDRT for more than three minutes could lead significant decline in situation awareness~\cite{de2014effects}. Thus, in this study, each NDRT lasted about 219 seconds~(SD=15s) before the system initiated a TOR. Participants interacted with each NDRT for about 657 seconds in each block of experiment. Both blocks of experiment consisted of the following NDRTs: 

\emph{\textbf{Watching}}. We selected two movies in the same Action/Thriller genre to prevent potential effects from one specific genre. Participants were given two Netflix movies to choose from, "Extraction" by Sam Hargrave or "Ava" by Tate Taylor. 

\emph{\textbf{Reading}}. "No One Is Too Small to Make a Difference" by Greta Thunberg was selected for the users to read. Participants were also instructed to read out loud to make sure they are surely reading the book.

\emph{\textbf{Conversing}}. The subjects were asked to have a conversation with the experimenter sitting behind them to simulate conversation with another passenger regarding everyday topics (e.g., plans for summer vacation).

\emph{\textbf{Gaming}}. The participant played a 2048 smartphone game, a single-player sliding block puzzle game, whose objective is to slide and combine numbers on a grid with the purpose of achieving a sum of 2048. This game challenge physical and visual demands for receiving an emergency alert.

%====================================
%====================================

\subsection{Procedure} ~\label{Sec:Procedure}
Upon arrival, the participants were briefed about the study. Participants then signed an informed consent form and completed a demographics questionnaire, followed by a 5-minute practice drive to get familiar with the driving simulator and NDRTs. 
We fitted the participant with the Shimmer3+ wearable device and calibrated the eye-tracker algorithm~(which was re-calibrated at the beginning of each trial). Participants were informed that there was no need to actively monitor the driving environments or resume the control of the vehicle unless a TOR was issued. However, they were instructed to resume the vehicle control as soon as a TOR was issued, then switch back to the automated driving once the incident had passed and continue the engagement with a NDRT. 

At the beginning of the drive, the participants were asked to activate the automated mode and perform a NDRT based on the experimenter's instructions, followed by three more NDRTs (see \tabref{tab:NDRTs}). Previous research finds that participants engaging with a NDRT for more than 180~seconds could lead to a significant decline in situation awareness~\cite{de2014effects}. 
In this study, immersion to a NDRT lasted 200~seconds on average (SD=15s) before being interrupted by a TOR, which was programmed to be triggered automatically about 111~meters~($\approx$5s) before detection of a dangerous incident. The advisory warnings were also triggered 38-45~seconds (M=40.3s, SD=1.6s) prior TOR to make drivers vigilant of vehicle's state. Overall, participants engaged with each NDRT per trial for 12-15min. The chosen time window is twice as long as in previous studies~\cite{van2017priming,borojeni2018reading} in order to evaluate CAWA's impact on driver takeover readiness. At the end of each trial, the questionnaire on workload~(DALI) and perceived safety and urgency were administered. 

After the participant completed all of the driving trials, the experimenter conducted a semi-structured interview to seek the participant's general feedback about the study. The interview guideline was prepared following a prior study~\cite{trosterer2017we}. 
The entire study took about 100-130~minutes, and the participant received a \$30 gift card for completing the study. 
%====================================
%====================================
\subsection{Dependent Variables}
To investigate the proposed research questions, we used the following objective measurements and subjective feedback as dependent variables. 

\textbf{RQ1} questions driver takeover behavior. We measured the driver's reaction time (i.e., the time difference between the TOR initiation and the exact moment of the driver pressing the button on the steering wheel to resume manual control), and the lateral vehicle control (i.e., deviation from the lane during the takeover). 

\textbf{RQ2} asks about driver situational awareness. As gaze behavior shown to be a reliable indicator of situation awareness~\cite{bhavsar2017quantifying,recarte2000effects,li2012evaluation}, we applied the state-of-the-art computer vision techniques~\cite{park2018learning,park2018deep} to estimate the gaze behavior of drivers in real-time. We calculated two metrics: (i) percentage of drivers looking at the road; and (ii) fixation duration of when a driver's eyes are on/off the road. 

\textbf{RQ3} evaluates driver stress and cognitive workload. We used the biometric data to calculate metrics including heart rate variability and the number of GSR signal peaks, showing mental workload and stress respectively.pNN50 was calculated as the number of two consecutive intervals (called NN) in which the change in consecutive normal sinus intervals
exceeds 50 milliseconds divided by the total number of NN intervals measured.
Furthermore, we report the number GSR peaks from the time of advisory warning receipt to moment of takeover control.
We also asked participants to complete the Driving Activity Load Index~(DALI)~\cite{pauzie2008method}, which customizes NASA-TLX for the automotive domain. 

\textbf{RQ4} inquires about driver perceptions. We asked participants to rate their perceived safety, disruptiveness, and the urgency of advisory warnings on a 5-point Likert-type scale ranging from 1 (strongly disagree) to 5 (strongly agree), which was adapted from the rating questionnaire used in the prior study by Iqbal et al.~\cite{iqbal2011hang}. At the end of the study, we interviewed the participants
about their preferences for the different advisory warnings and solicited their rationales for the order of preference and usefulness.
% \vspace{-0.2in}

%====================================

\section{Results} \label{sec:results}
\pdfoutput=1
We analyzed the data collected from the user study for the proposed research questions. 
We set the statistical significance level as $\alpha = 0.05$. 

%======================================================
\subsection{Quantitative Measurements}
\subsubsection{Effects on Driver Takeover Behavior (RQ1)}

We observed in the study that participants were able to take over the vehicle control following TORs with a high success rate. Out of the 456 TORs (19 participants $\times$ 2 trials $\times$ 12 true TORs per trial), only 4 takeovers were failed (e.g., the driver was playing a game on the mobile phone and failed to take over in a timely manner, causing the vehicle to collide with an obstacle).
We conducted statistical analysis using the data of 452 successful takeovers to investigate drivers' takeover behavior.

% \textcolor{red}{add the values for CAWA vs Baseline...Post-hoc Bonferroni tests }
% We found significant main effect of NDRTs~($F(3,443)=2.39$, $p=0.034$, $\eta^2=0.049$), and type of advisory warnings~($F(1,443)=185.53$, $p<0.001$, $\eta^2=0.47$) on reaction time~(see Fig.\ref{fig:Rt_box}). But, no interaction effect was found~($F(3,443)= 1.42$, $p=0.21$, $\eta^2=0.01$). Post-hoc analyses with Bonferroni revealed that there was a significant difference between gaming on the phone and conversation with the experimenter~($p<0.01$).It illustrates that drivers could see an upcoming potential hazard required an early action. It also indicated a significant difference between gaming and watching a movie on tablet~($p<0.01$).But, there was no significant difference between other activities. 

\begin{figure*}[t!]
\centering
  \includegraphics[width=0.67\textwidth]{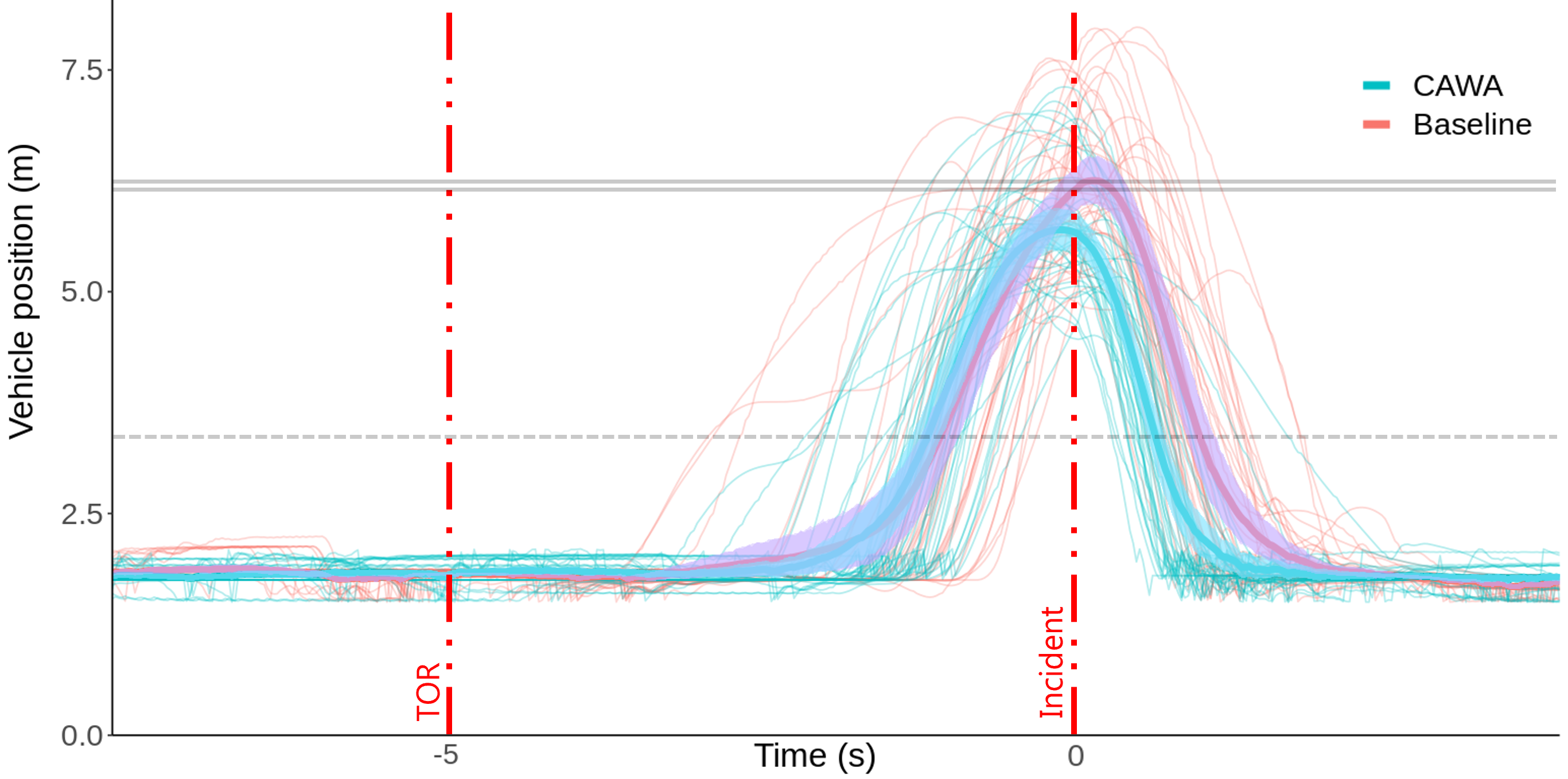}
    \vspace{-0.1in}
  \caption{Lateral trajectories of vehicle after TORs.}
  \label{fig:trajectory}
\end{figure*}

\textit{\textbf{Takeover Quality}}. We plotted the vehicle trajectories in \figref{fig:trajectory}. It shows substantial variation in control strategies and higher takeover control after receiving CAWA, as opposed to the baseline, indicating better takeover quality.

A two-way repeated-measures ANOVA also found statistically significant effects on the lateral vehicle control ($F(1,443)=13.46$, $p<0.01$, $\eta^2=0.15$) by comparing CAWA and the baseline. Post-hoc showed that the \textit{visual} warning resulted in lower lateral deviation compared to all other modalities~($p<0.01$). This means that the drivers who were looking at the road while holding a conversation had better control of the car as opposed to other modalities.

%--------------------
\begin{figure*}[ht!]
    \centering
    \includegraphics[width=0.97\textwidth]{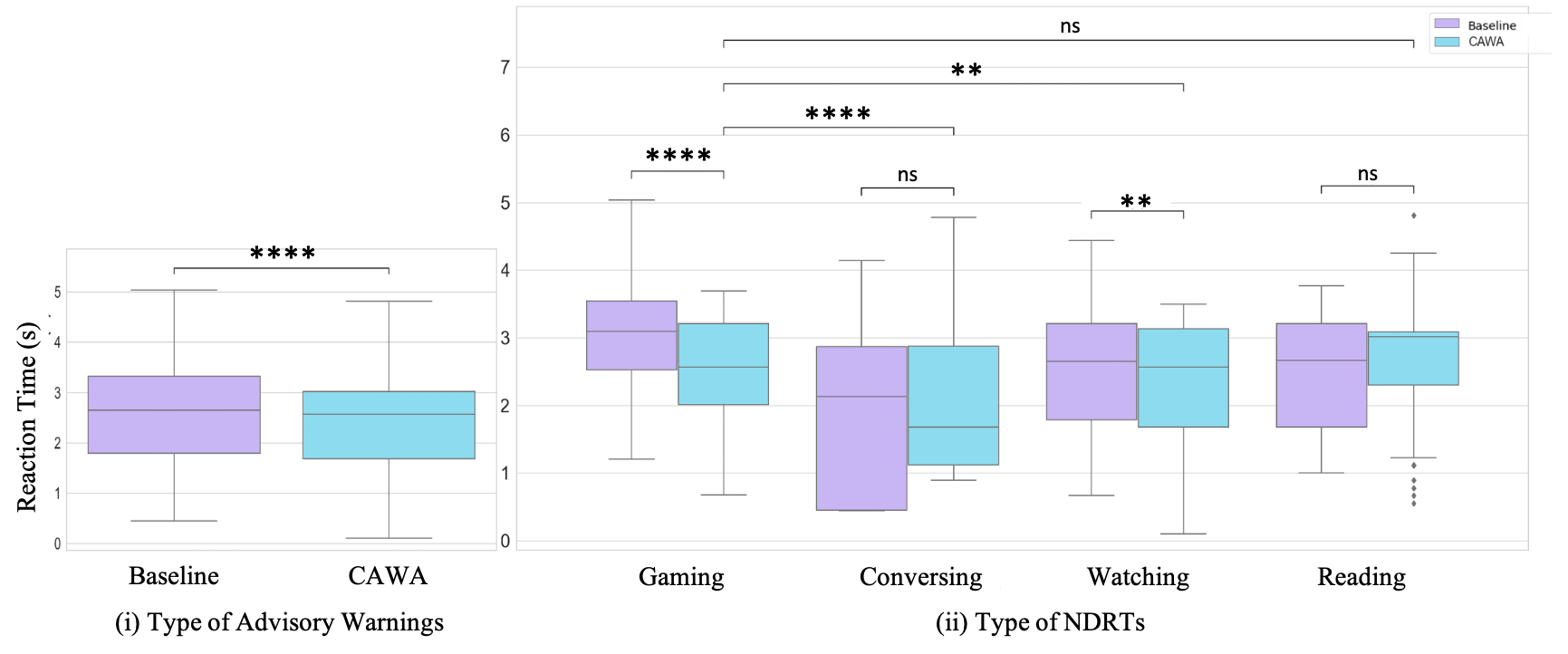}
  \caption{Comparisons between the participants’ takeover reaction time in relation to the type of advisory warning and the imposed modality. **: $p<0.01$, ****:$p<0.0001$ }
  \label{fig:Rt_box}
\end{figure*} 

%--------------------
% \vspace{-0.09in}
%--------------------

\textit{\textbf{Reaction Time}}. A two-way repeated-measures analysis of variance (ANOVA) analysis found a significant main effect of type of NDRTs~($F(3,443)=2.39$, $p<0.05$, $\eta^2=0.049$) and type of advisory warnings~($F(1,443)=185.53$, $p<0.001$, $\eta^2=0.47$) on reaction time, showing CAWA can lead to a faster reaction time than the baseline.
For types of NDRTs, post-hoc analyses with Bonferroni revealed that there was a significant
difference between gaming on the phone an conversation with the experimenter~($p<0.01$) and between gaming and
watching a movie on tablet~($p<0.01$), indicating that conversing with passengers and watching movie leads to quicker reaction time than gaming (see Figure~\ref{fig:Rt_box} (ii)).

%--------------------
\begin{figure}[t!]
\centering
  \includegraphics[width=0.98\linewidth]{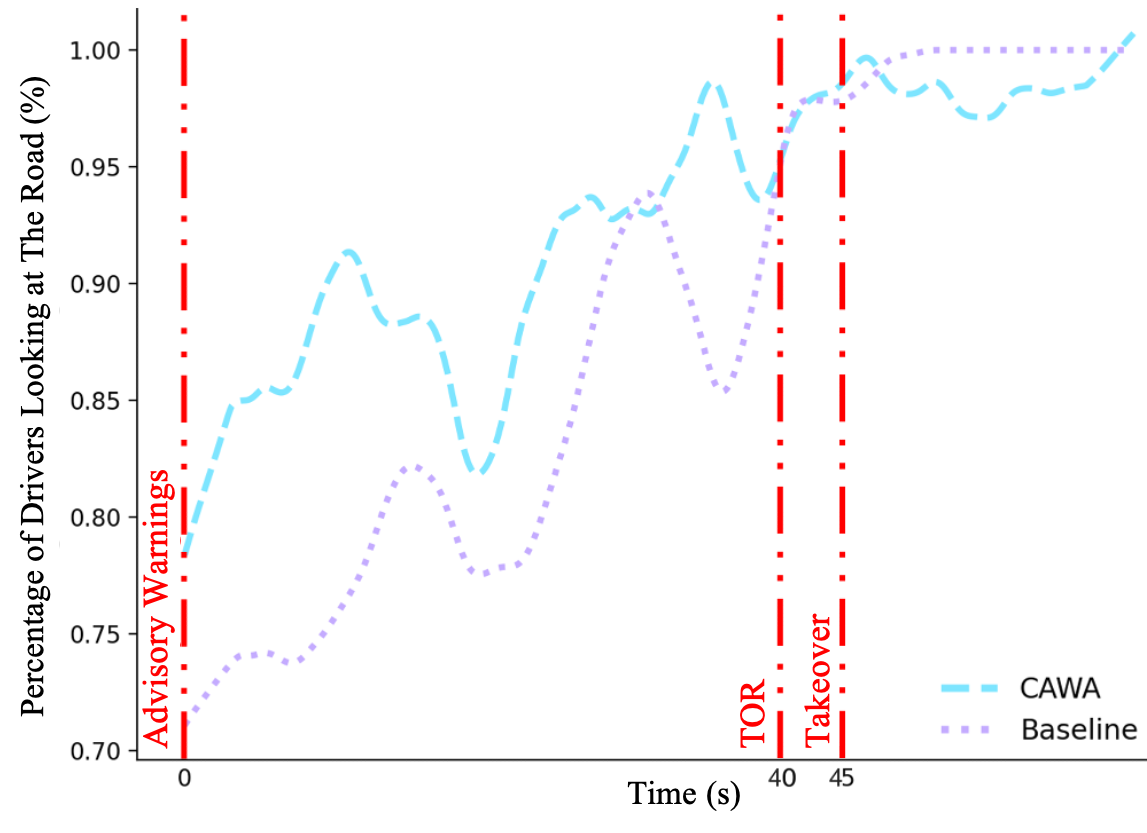}
  \caption{Results on the percentage of drivers looking at the road.TOR: issue of TOR; Takeover: the longest time of takeover}
  \label{fig:gaze_percentage}
\end{figure} 
%--------------------
% \vspace{-0.6cm}

%======================================================
\subsubsection{Effects on Driver Situational Awareness (RQ2)}
\figref{fig:gaze_percentage} displays the percentage of drivers looking at the road from the time they received the advisory warning to 20 seconds after resuming vehicle control~(i.e., the number of drivers looking at the road at a given time divides the total number of participants). On average, 87.6\% of the drivers look at the road from the time of receiving an advisory warning, to the time of actual takeover of control, showing an enhancement on driver's situation awareness. Shortly after the TOR, more than 95\% of drivers shifted their visual attention to the screen. However, more participants stayed vigilant in baseline after taking the vehicle control.
Furthermore, analyzing the eye-gaze vector for investigating the fixation time on/off the road, shows the standard deviations across the mean of participants. We ran ANOVA and found significant main effect of type of advisory warnings~($F(1,443)=39.47$, $p<0.05$, $\eta^2=0.23$) on the fixation time. Although conversing resulted in higher time fixation on the road, there was no significant difference was observed between the type of NDRTs.

%======================================================

\subsubsection{Effects on Driver Stress and Cognitive Workload (RQ3)}

% Prior research has found that heart rate variability (e.g., pNN50) and galvanic skin response (GSR) are good indicators of drivers' stress and cognitive states~\cite{munla2015driver,manawadu2018tactical}. 
We investigated the effect of CAWA and baseline on stress~(i.e. GSR) and cognitive load(i.e. heart rate variability(HRV)). The results show no significant effect of NDRT type~($F(3,443)=0.95$, $p=0.42$, $\eta^2=0.007$) and type of advisory warnings ($F(1,443)=2.23$, $p=0.14$, $\eta^2=0.006$) on HRV~(i.e., pNN50). Besides, the statistical analysis showed that the number of GSR peaks from the time of receiving advisory warnings to moment of takeover was significantly impacted by type of NDRT~($F(3,443)=0.95$, $p=0.42$, $\eta^2=0.007$), no significant effect of the type of advisory warnings was found~($F(1,443)=2.23$, $p=0.14$, $\eta^2=0.006$). Post-hoc test with Bonferroni on the number of GSR peaks indicated a statistically significant difference between watching a movie with conversing~($p < 0.05$) and reading~($p<0.05$).

% the statistical analysis showed that the number of GSR peaks from the time of receiving advisory warnings to the moment of takeover was significantly impacted by type of NDRTs~($F(3,443)=0.95$, $p=0.42$, $\eta^2=0.007$). However, no significant effect of the type of advisory warning was found~($F(1,443)=2.23$, $p=0.14$, $\eta^2=0.006$). Post-hoc test with Bonferroni on the number of GSR peaks indicated a statistical significance between watching a movie with conversing~($p < 0.05$) and reading~($p<0.05$). 
% One possibility for higher GSR peaks could the selected genre of the movies demonstrated an intense level of arousal associated with human emotions.

% \begin{figure*}[t!]
% \centering
%   \includegraphics[width=.8\textwidth]{Figures/demand_new.pdf}
%   \caption{Results on DALI ratings about workload.}
%   \label{fig:demand}
% \end{figure*} 
%--------------------

\begin{figure*}[t!]
\centering
% \hspace{-1cm}%
\includegraphics[width=\textwidth]{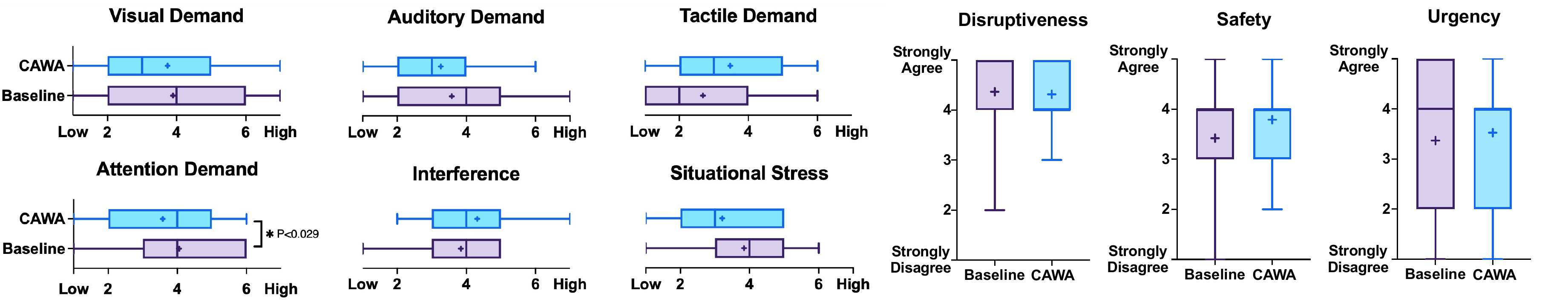}
\subfloat[\label{DALI} Results on DALI ratings]{\hspace{.5\linewidth}}
\subfloat[\label{safety} Results on driver perceived safety, disruptiveness, and urgency]{\hspace{.55\linewidth}}
\caption{The results of the (a) driving activity load index (DALI) questionnaire along with (b) perceived disruptiveness, safety, and urgency under two CAWA and Baseline conditions. \label{subjective}}
\end{figure*}

We also analyzed the participants' subjective ratings on DALI, which includes six dimensions of workload as shown in \figref{DALI}.
ANOVA analysis found significant effects on attention demand (F(2,54)= 3.70, $p<0.05$, $\eta^2= 0.12$). Post-hoc testing with Bonferroni on attention demand also indicated a significant difference between CAWA and the baseline~($p=0.029$), which means that the attention required by the baseline was much more demanding than CAWA. However, no statistically significant effects were found in other workload dimensions.

%========================================================

%--------------------
% \begin{figure*}[t!]
% \centering
% % \vspace{-1cm}
%   \includegraphics[width=\textwidth]{Figures/combine3.pdf}
%   \begin{minipage}{.55\textwidth}
%         \centering
%         \vspace{-0.15in}
%         \captionof{figure}{Results on DALI ratings about workload.}
%           \label{fig:demand}
%           \end{minipage}%
%   \begin{minipage}{.4\textwidth}
%         \centering
%         \vspace{-0.15in}
%       \captionof{figure}{Results on driver perceived safety, disruptiveness, and urgency of advisory warnings.}
%     \label{fig:subjective1}
%           \end{minipage}
%   \vspace{-0.15in}
% \end{figure*}

%--------------------

\subsubsection{Driver Perceptions (RQ4)}

%--------------------
% \begin{figure*}[t!]
% \centering
%   \includegraphics[width=.8\textwidth]{Figures/subjectivenew3.pdf}
%   \caption{Results on driver perceived safety, distruptiveness, and urgency of advisory warnings.}
%   \label{fig:subjective1}
% \end{figure*} 
% \begin{figure*}[t!]
% \centering
%   \includegraphics[width=1.05\textwidth]{Figures/subjective3.pdf}
%   \caption{Results on driver perceptions.}
%   \label{fig:subjective1}
% \end{figure*} 
%--------------------

\figref{safety} shows the survey results on drivers' perceived safety, disruptiveness and urgency of advisory warnings. The results of safety~($F(2,54)=0.799$, $p=0.377$, $\eta^2=0.021$), disruptiveness~($F(2,54)=0.0.498$, $p=0.485
R$, $\eta^2=0.014$) and urgency~($F(2,54)=2.866$, $p=0.099$, $\eta^2=0.074$), did not show a significant main effect on type of advisory warnings. Even though more participants rated CAWA to be safer with higher urgency than the baseline, yet they found it more disruptive.

\subsection{Qualitative Measurements}
\subsubsection{Preferences and Challenges}
% More participants strongly agreed that CAWA was safe, while fewer of them strongly agreed that CAWA was disruptive and urgent. 
% \textcolor{red}{In the post-session interview, the majority of participants (18 out of 19) expressed more positive feelings toward CAWA and found it more gentle. However, one participant, who was uncomfortable with vibrotactile advisory warnings, declared otherwise.}
For the qualitative evaluation, the details of the interviews for each subject were recorded verbatim. We transcribed the audio recordings from the post-session semi-structured interviews into text and arranged the texts according to the condition. Then, based on the participants’ statements on each condition describing their observation, we compared the similarities and differences. Overall, seventeen participants rated the CAWA as more gentle than baseline warnings. Two participants perceived baseline as more gentle mainly due to the ``shocking'' of the \textit{Vibrotactile} modality.
%advisory warnings .  
CAWA was referred to as “safer” alternative by fifteen participants. 

They described a feeling of a need for learning why they received the warnings in order to adapt to the situation. However, four participants found CAWA “disruptive'', “too pressuring”, and “urgent''. Although they stated that only a beep is “not enough” or there is “not enough information”, they still preferred being less disturbed and let continuing engagement in NDRTs.

% \subsubsection{Types of warnings} 
\subsubsection{Types of Modalities}
Participants were asked to express their perception about pros and cons of each implemented type of warning modalities, and their preferences were varied. Participants’ preferences of the most suitable types of warning modality for increasing situation awareness and takeover readiness were ranked as \textit{Text messages} (N = 7), \textit{speech-based} (N = 6) and \textit{Vibrotactile} (N = 4), \textit{Visual} (N = 1). Only one participant mentioned that he doesn't need any warnings at all.
Four participants expressed the main reasons for preferring the \textit{vibrotactile} modality over the others was as ``\textit{it directly connected to my body and waked me up}'' and they preferred feeling the cues rather than being interrupted via visual or auditory alarms (P15). For example, a participant stated ``\textit{... I guess if you are in the car and your have your music up really loud and watching TV really loud then vibrotactile warnings would be really helpful''}. 
%On the hand, 
However, five participants did not favor the \textit{vibrotactile} modality found it difficult to know where their attention should be directed to, for example, P2 stated - ``\textit{it did not vibrate anywhere I need to pay attention or I was close to accident. I don't know in which condition did it vibrate or what I should do''}. 
% The only condition, may it helps would be when I was sleeping, otherwise I didn't find it useful}'' 
% Several participants who dislike the textile feels that the textile is not application-specific, only when they are feeling drowsy, for example sleeping, for example, 
% One participant found the textile feedback confusing as it directs his attention to nowhere, as P1 stated ``For other context-aware warnings, I know where to look at ''  
% The majority (two-thirds) of participants found the textile notification as the most gentle and potentially useful way as an input of notification,  

Seven subject found the text messages notification “very useful'', “creative'', “attention-grabbing'' while engaging with the games on the cellphone. For example, P5 stated - \textit{``The game was the hardest. With the game I was using my hands and my eyes, and then when the computer says takeover, I need to redirect my eyes to the screen, and put down the phone and then hit the button, with the book I can quickly put it down, and then put it back}''. However, five of participants who opposing the text messages mentioned two main reasons. They found it “disruptive'' as it could block the other urgent text notifications during everyday life. In addition, it was expressed by one that the workload that they need to not only pay attention but also read the text message.

Participants had mixed feelings over the
\textit{Speech-based} modality, as they perceived it as most interruptive and ``jarring'' of the four, yet effective; Six participants valued it as ``\textit{it stands out from everything else, and immediately brought me back.} Contrarily, over half of participants perceived the speech-based modality as ``robotic''. 
% Several participants mentioned that hearing the robotic voice perceived as jarring that they may paying more attention and felt more urgent. ``\textit{A warning can fade into the background when you were doing the task, but you are always able to hear the voice, all of the games, the voice was way louder than any others, it stands out from everything else, and immediately brought me back.}'' However, participants disliked the voice notifications felt the voice interruptive and sounds similar to takeover request that they sometimes over-responded to it ``\textit{the voice was shocking and startling, i was nearly jumped out when i was playing, and also it sounded very similar to the takeover request.}''

Three participants favored the \textit{Visual} modality, as the most “practical'' type of warning. These participants backed their choice as it required "less attention" and it was found "less annoying". For instance, P15 stated that \textit{Visuals} is "\textit{easy to understand compared to the text messages that I still need to read the words.}" Three participants opposed visual warnings as it potentially ``\textit{occluded the vision of the situation}'' (P1, P5, P7) and could be ``distracting'' (P7). For example, P5 commented - ``\textit{it can blend into the background''}.

\section{Discussion} \label{sec:discussion}
This study aimed to investigate the effects of context-aware advisory warnings on takeover readiness and performance. In order to do so, we proposed a novel context-aware advisory warning system~(CAWA). CAWA adapts its warning modalities based on the context a driver is immersed in. In contrast to pre-alert systems~\cite{van2017priming} that startle and stress the driver to take an immediate action, advisory warnings are non-assertive. Although a large body of literature has investigated the influence of various warnings on takeover time~\cite{lu2017much,eriksson2017takeover,petermeijer2017driver} and quality~\cite{du2020evaluating,weaver2020systematic}, to the best of our knowledge, it is the first study to employ multiple modalities for “advising'' drivers of automated vehicles to pay attention to the driving scene and to be more conscious of the automated driving status, specifically via text messages.

% Some studies on transitioning the vehicle control in automated driving have suggested that cognitive state of drivers are more important than  motor readiness when designing an effective TOR~\textcolor{red}{\cite{}}, while other studies have suggested that the physical attributes that are captivated by NDRTs are essential~\textcolor{red}{\cite{}}. Therefore, we considered both aspects in our detailed assessments.
% We evaluated takeover reaction time, takeover quality, situational awareness, perceived safety and urgency of drivers after receiving CAWA along with an auditory warning. 

% Results revealed the importance of advisory warnings for driver's readiness and situation awareness.
% Results showed a significant differences between CAWA and baseline in controlling the vehicle at moment of takeover. There were also significant differences in reaction time due to increases in fundamental frequency. In terms of situation awareness, 
% while more drivers tend to look at the screen longer after receiving CAWA, there were no significant difference. DALI also revealed a significant difference on attention demand. The overall results show that divers received CAWA have higher takeover quality and lower reaction time when compared to drivers received auditory warning.

% However, the user experience of the proposed method was disparate. Since the speech-based auditory is an already applied modality for communicating with drivers, we compared CAWA with it to investigate the effectiveness of CAWA on preparing drivers for better takeover performance and higher situation awareness.

\subsection{Takeover Behavior}
Takeover reaction times and quality were measured and analyzed to compare differences due to perceived CAWA and auditory warning. In line with previous studies that found auditory warning leads to significantly higher reaction time~\cite{eriksson2017takeover,politis2015language}, we observed significantly higher reaction times with baseline as opposed to CAWA. Further, the  results showed that conversing yielded the lowest reaction time, but the results may reflect the fact that the conversation with the experimenter did not need shifting visual attention. The most cognitively and visually demanding task, playing 2048 game, showed higher reaction time. Although the react times were varied, CAWA helped drivers to resume the control faster. The range of reaction time obtained in our study slightly differ from previous studies~\cite{eriksson2017takeover,zhang2019determinants}, showing that participants were somewhat prepared to take the control or anticipated a takeover after receiving an advisory warning. 
Despite the research of~\cite{gold2018modeling} indicating that the complexity of NDRTs is not a significant variable for reaction time, our experiment's findings indicated that takeover time was significantly impacted by physical and cognitive loads needed for performing NDRTs. 

We also observed that CAWA assisted drivers to departure earlier and helped less deviation from the center of the lane(see Figure~\ref{fig:trajectory}). 
This finding of vehicle control after receiving TOR is in line with our expectations based on previous studies~\cite{politis2015language,manawadu2018tactical} showing non-auditory warnings provides relatively better control of the vehicle.  Our findings also suggest that a safer takeover is a composite of multiple factors~(e.g. type of NDRT and its level of complexity, type of modalities, etc.) and they may have a greater effect on readiness and takeover.

\subsection{Situation Awareness}
We observed higher rates of monitoring of the road after receiving CAWA compared to the baseline. More specifically, after the vehicle approached to advisory warning time, 14\% more of driver looked back at the road and stayed more visually attentive. In general, our results shows that receiving advisory warnings increases 26\% likelihood of looking at the road as opposed to the results reported in~\cite{van2017priming}. 

\subsection{User Experience}
Concerning the usability aspect of proposed method, the users perceptions towards advisory warnings' safety and disturbance were analyzed along with their subjective workload using DALI survey. Participants’ ratings of their perceived safety, disruptiveness and urgency, favored CAWA, but did not differ significantly between the two conditions. Post-study interviews revealed that users believed that CAWA could avoid being missed, but it leads to higher annoyance. Even though we extended the timing of advisory warning suggested by literature to 200s on average, we acknowledge that a better experimental design with less frequent interruption could have increased CAWA's usability. In addition to driver's perceptions, only the significant difference in the attention demand subscale of the DALI supported the hypotheses. Despite slightly better score in visual and auditory demand of CAWA, participants' subjective workload rating did not differ significantly between the conditions. It is possible that the similar time budget to takeover between the two conditions was perceived as alike workload. Another possibility for the absence of significance in the subscales of DALI could be due to the within-subject design where we only collected one data point to compare the conditions.

% Additionally, the interview showed that the least preferred modality was always visual. These results contradict previous findings~\cite{dass2013haptic} indicating that users did not prefer vibrotactile over visual. 
% That is, drivers of automated driving would rather to do not be interrupted and stay immersed in the activities.

\section{Limitations and Future Work} \label{sec:limitations}
We applied unimodal advisory warning rather than multimodal modalities. While multimodal modalities were found to improve reaction time~\cite{petermeijer2017driver} and quality of takeover~\cite{naujoks2014effect}, prior studies reported them as urgent~\cite{kutchek2019takeover} and annoying~\cite{politis2015language}. We utilized unimodal modalities (1) to avoid resource sharing conflicts according to Wickens’ multiple resource theory~\cite{wickens2002multiple}, (2)to investigate the impact of non-assertive advisory warnings on takeover behavior. However, we acknowledge that a more exhaustive picture would have been available if we combined multiple modalities to urge drivers to pay attention to the driving scene. 

% In accordance with our desire to design advisory warnings (rather than takeover requests) to keep drivers vigilant when dong NDRTs, we aim to trade-off the driver readiness (less urgency) and effectiveness (more urgency) in CAWA deign. Therefore, the main design goal of CAWA is to \textit{balance} the underlying cognitive load and utility of enhancing driver vigilance before TORs. Furthermore, we only compared it to the non-context aware warning (speech-based warning) in the study, yet future work can consider comparing multiple modalities context-aware warnings to CAWA to investigate the potential pros and cons.}
% One shortcoming of the study is that no multimodal modality was employed to warn drivers. Although the literature suggests employing a multimodal warning system to improve reaction time~\cite{petermeijer2017driver} and quality of takeover~\cite{naujoks2014effect}, we intentionally utilized unimodal warning system in this study for two main reasons. (1) To avoid resource sharing conflicts according to Wickens’ multiple resource theory~\cite{wickens2002multiple}; (2) To investigate the impact of CAWA on takeover behavior and situation awareness. 
% A more exhaustive picture would have been available if we combined multiple modalities to urge drivers to pay attention to the driving scene. 
    % \item 
Another limitation is using a driving simulator. While driving simulator studies are very common due to advantages in creating standardized situations for experimental control, they come with limited external validity. Participants may react differently in the lab than they do naturally while driving in the wild. Despite randomizing the time interval for advisory warnings, participant could still expect to encounter a TOR.
% \item In addition, our study participants were all students recruited from the anonymous university. Thus, it is unknown if our findings can be extrapolated to different age-groups, social norms, and cultural values. 

% \end{itemize}
Despite these limitations, this study takes the first steps toward enabling CAWA for automated driving, which can provoke many exciting future research directions. In this study warnings were triggered for a fixed period (about 40 seconds) before TORs in the study. Future work could leverage recent advances in predicting driver takeover behavior and readiness~\cite{yoon2021modeling,pakdamanian2021deeptake}, and develop agent-based systems to intelligently decide when and how to trigger warnings based on driver state predictions.

\section{Conclusion} \label{sec:conclu}
In this work, we proposed CAWA, a novel method that provides gentle advisory warnings to improve driver readiness. Furthermore, CAWA tends to select appropriate modality according to the context of NDRTs, seeking the balance between (i) avoiding a warning to go unnoticed like generic warnings and (ii) frequent interruptions unless situation awareness falls bellow dangerous level for proper takeover in the case of emergency. Our user study found encouraging results proving the applicability of potentially incorporating CAWA into the design of automated driving vehicles for safer and smoother transition of control.

% effectiveness of CAWA, in terms of safer takeover behavior, improved driver situational awareness, and less attention demand.% and more positive user feedback.

\section{ACKNOWLEDGMENT}
This work was supported in part by National Science Foundation (NSF) grants CCF-1942836 and CNS-1755784.
Any opinions, findings, and conclusions or recommendations expressed in this material are those of the author(s) and do not necessarily reflect the views of the grant sponsors.

%% The next two lines define the bibliography style to be used, and
%% the bibliography file.
\bibliographystyle{ACM-Reference-Format}
% \begin{thebibliography}{}
%   \input{Sections/references}
% \end{thebibliography}
\bibliography{Sections/references}

\end{document}